# Breaking the Myth of Four-Tier Stacked Transistors for Logic: an Investigation from Process to Device and Block-Level PPA

Jianxiang Jin[1], Haoran Lu[1], Yihan Luo[1], Yanbang Chu[1], Yibo Zhang[1], Jingru Jiang[1], Yuji Cheng[1], Wanyue Peng[1], Xun Jiang[1], Feiyu Teng[1], Kairang Guo[1], Ming Li[1], Yibo Lin[1], Runsheng Wang[1], Heng Wu[1,*], Ru Huang[1]
[1]School of Integrated Circuits, Peking University, Beijing, China, *Email: hengwu@pku.edu.cn

***Abstract*****—**For the first time, this work evaluated whether 4-tier stacked transistors can translate the straightforward density gains into realistic chip-level scaling benefits. A practical dual-sided 4-tier transistor-stacking CFFET (Flipped CFET-to-CFET) process was proposed, on which a DTCO framework spanning standard-cell design, local interconnects, RO analysis, and physical implementation was built under 2A node design rules. The CFFET was also benchmarked with its 2-tier transistor-stacking counterparts FFET. Among CFFET configurations, the best-performing Symmetric Power still lags FFET by 11.2% in RO frequency due to the complex MOL parasitics. At the block level, CFFET reduces the Ibex RISC-V core area by only 12.8% (far below the 40% nominal cell-height reduction), while running 10.9% slower at iso-power. We further introduced a novel Tetris-like cell stitching to minimize the area overhead, which allows non-rectangular cell design and recovers 7.4% area in a 1-bit ALU design. An optimized NanoStack CFFET was finally proposed, inspired by previous work. This work indicates that straightforward vertical 4-tier stacking proves insufficient to guarantee enough PPA benefits, though the resulting design insights may guide future advanced architectures.

## I. Introduction

As conventional dimensional scaling slows, stacked transistors such as CFET [1-6] and FFET [7,8] have attracted significant attention as scaling options beyond 1nm (Fig. 1). Multi-tier stacked logic transistors are also being widely discussed [7,9,10], trying to replicate the approaches in 3D-NAND [11] or 3D-DRAM [12]. However, for logic applications, more transistor-stacking tiers may not necessarily yield practical PPA benefits, as standard cells suffer from irregular routing, stringent pin-access requirements, and strong device-to-interconnect coupling. Thus, as a representative case, 4-tier transistor architecture must be evaluated through standard-cell, circuit, and block-level implementation.

This Design-Technology Co-Optimization (DTCO) research presents a thorough study on the 4-tier transistor-stacking **CFFET**, with vertically back-to-back stacked CFETs and FFET-like dual-sided interconnects. By forming CFET on each side of the wafer separately with self-aligned active, the CFFET process greatly reduces high aspect-ratio (AR) steps (dummy gate, S/D epi, HKMG) in monolithic CFET-on-CFET flow. Power and FET polarity configurations of CFFET were first studied against the 2-tier transistor-stacking FFET under identical 2A design rules. Though Symmetric Power with N-P-P-N transistor stack outperforms other CFFET configurations, it still lags FFET due to increased MOL parasitics. It was further investigated on an Ibex RISC-V core, revealing limited area gain and notable performance loss from local-interconnect parasitics. To mitigate the area loss imposed by the 4-transistor cell building block, a novel **Tetris-like cell stitching strategy** was introduced, supporting flexible, non-rectangular multi-tier cell designs and post-placement EDA integration. However, it recovers only part of the lost area and does not address the MOL parasitics responsible for performance degradation. What's more, CFFET SRAM with extreme single-row bit cell design was explored. Finally, the extendibility of CFFET was examined and a novel **NanoStack CFFET** was proposed. It shows better routing flexibility and lower parasitics by offsetting active of FS/BS CFETs, resulting in 3.1% higher RO frequency than FFET. However, the greatly increased process complexity requires further investigation rather than simple stacking.

## II. Process and Cell-level Assessment

***A. Fabrication of CFFET*** Fig. 2 illustrates the key process flow of CFFET. All 4 tiers' actives are first patterned together, following an approach similar to FFET [7,8]. A common-gate (CG) monolithic CFET with Vertical-Horizontal-Vertical (VHV) routing and BEOL stack is then formed on the wafer frontside (FS). After wafer flipping and backside thinning, the STI covering the backside (BS) active fins is etched down, where a BS CG CFET is formed. Note that the NFET/PFET arrangement within each CFET remains configurable. Compared with the fully monolithic CFET-on-CFET process [10], this dual-sided flow distributes the four transistors between FS and BS CFETs. It therefore reduces challenges in vertical patterning with extremely high AR, such as dummy gate, S/D epi formation and HKMG where a monolithic 4-tier common gate could reach 370 nm in height with severe fabrication difficulties, offering a more practical flow toward 4-tier integration. Inter-tier vertical access, however, still relies on MOL supervias and VHV routing for M0 access to bottom transistors (Fig. 3), requiring highly selective dielectric etching at AR up to ~40, which remains a key process concern and further induces layout complexity and parasitic penalties in standard-cell design.

***B. TCAD and Compact Model*** Since this research mainly focused on the interconnect and layout, all NFETs/PFETs were each assumed to have identical intrinsic device characteristics regardless of their tier position. All devices used for the ring oscillator (RO) and library characterization have fixed $I_{off}$ of 2 nA [13]. BSIM-CMG compact models (CMs) [14] were fitted to TCAD $I_dV_g$, $I_dV_d$ and CV curves**.**

***C. CFFET Baseline Configuration Selection*** The power and FET-polarity configurations of CFFET were first benchmarked to identify the best-performing baseline for subsequent evaluations. Three power configurations were compared against the 2-tier transistor-stacking FFET (cell height = 50 nm, either VDD or VSS to each wafer side) under the same 2A design rules (CPP = 45 nm): BS Power [10], Staggered Power [15], and Symmetric Power [7], representing progressively improved FS/BS symmetry (Fig. 4). Circuit-level performance was evaluated with 15-stage INVD2- and NAND2D1-based ROs with fan-out of 3. For INV, Symmetric Power achieves the highest frequency at iso-power without BEOL load by balancing and reducing source resistance to power. BS Power and Staggered Power are respectively 9.0% and 6.3% slower @ $V_{dd}$ = 0.7 V (Fig. 5(a)). Even so, the best CFFET configuration remains 11.2% slower than FFET, and the gap widens to 12.4% for NAND2D1, whose more complex MOL interconnects lead to higher parasitic RC. Parasitic partition of INVD2 (Fig. 6) shows that inter-tier MOL supervias and VHV routing degrade the output/source resistance and the effective output capacitance over FFET, indicating the worse performance is intrinsic to the 4-tier vertical-access topology. In addition to the performance penalty, power configurations also affect standard-cell routability. The two asymmetric power configurations block supervia sites and congest MOL routing in AOI22D2, whereas Symmetric Power balances power/signal access (Fig. 7). Table 2 summarizes the design flexibility of the three configurations.

With Symmetric Power fixed, FET polarity was further optimized between N-P-P-N and N-P-N-P (top to bottom); N–N–P–P was excluded as it requires split-gate patterning and cross-tier gate merging, unrealistic for practical implementation. Evaluated on INVD2 @ 0.7 V (4 GHz input, 10 ps transition, 1 fF load; Fig. 8), N-P-P-N exhibits better FS/BS transition-time matching and smaller worst-case rise/fall delay owing to its symmetric FS/BS M0 access.

Therefore, the **Symmetric Power with N-P-P-N stack** was identified as the best CFFET configuration and adopted as the baseline for subsequent block-level evaluations.

## III. Block-level PPA Scaling

By capturing design-closure and interconnect overheads, block-level implementation enables a more realistic assessment of PPA gains. Physical implementation (including dual-sided place-and-route) and PPA evaluation [16] were carried out on an Ibex RISC-V core using matched FFET and CFFET libraries with identical BEOL rules and physical-design settings (Fig 9(a)); post-route core layouts are shown in Fig. 9(b). Compared with FFET, CFFET reduces the routed core area by only 12.8%, closely matching the 13.5% instance-weighted cell-area reduction (Fig. 9(c) and 10(a)). Both values, however, remain far below the nominal 40% upper bound implied by effective cell height scaling from 50 nm in FFET to 30 nm in CFFET (60 nm/2). This shortfall originates at cell level, where MOL congestion requires additional CPPs and 4-transistor building-block induces transistor underutilization in certain cells. At 70% utilization, CFFET operates 10.9% slower than FFET at iso-power (Fig. 10(b)). Power breakdown at 1.5 GHz target frequency shows 2.3% lower net switching power in CFFET due to smaller core area, but 9.8% higher internal power from larger internal parasitics, leading to a 3.8% increase in total power (Fig. 10(c)). Thus, the 4-tier architecture doesn't provide a simultaneous PPA improvement: its limited area benefit is obtained at the cost of both performance and power.

Transistor underutilization is a unique challenge to 4-tier transistor-stacking CFFET, limiting area scaling: standard cell with an odd number of transistor pairs intrinsically leaves one 4-transistor column only partially occupied. To recover the area overhead, a novel **Tetris-like cell stitching strategy** was introduced: unused boundary transistors are released, yielding L-shaped cross sections that could be stitched with compatible neighbors (Fig. 11). In a proof-of-concept 11-gate 1-bit ALU (Fig. 12(a)), three stitching opportunities are identified, saving 4 CPPs in total cell width and achieving a 7.4% area reduction. In the Ibex netlist, stitchable cells account for 20% of all standard cells; however, each stitching pair saves only ~1-2 CPPs with the current CFFET cell design, bounding the projected block-level benefit at ~3%. The cell-stitching concept can also be extended to enable flexible non-rectangular multi-tier cell design, allowing each cell to select any transistor-tier pair and realize intra-cell routing through vertical access structures and M0 interconnects (Fig. 12(b)). From an EDA perspective, cell stitching can be integrated as a post-placement step: eligible cell pairs are identified and stitched before legalization, optimization, and routing. Cell stitching therefore confirms transistor underutilization as a source of the lost area gain but cannot close the ideal-to-realistic scaling gap. Overall, these results confirm transistor underutilization as one source of the lost area gain, and cell stitching alone cannot close the ideal-to-realized scaling gap. Nor is it expected to effectively mitigate the performance degradation, due to the underlying MOL parasitics.

## IV. SRAM Scaling for CFFET

4-tier transistor SRAM was further evaluated to assess the density benefit from structural regularity. Besides placing bit cells separately on FS/BS (2-tier SRAM, Fig. 13) [7], an extremely compact single-row CFFET bit cell design was proposed (Fig. 14(a)). Both achieved an equivalent 31.4% area reduction versus FFET SRAM under the same design rules (Fig. 14(b)).

Owing to the P-type pass-gate required by process assumptions, both stacked SRAM implementations exhibit improved RSNM but degraded WSNM than FFET (Fig. 15(a)). The delay degradation, by contrast, is dominated by interconnect parasitics extracted from the worst case bit cell in a 256×256 array. Although 2-tier SRAM reduces BL resistance ($R_{BL}$) through a larger M0 cross-section and CFFET SRAM further lowers WL resistance ($R_{WL}$) by reducing the cell height, the closer BL spacing and stronger WL-to-MD/gate coupling increase BL and WL capacitances ($C_{BL}$, $C_{WL}$), outweighing the resistance reduction benefit (Fig. 15(b)). In addition, the elongated cross-coupled feedback paths introduce extra storage-node resistance. Consequently, both 2-tier SRAM and CFFET SRAM exhibit degraded read/write performance despite the partial $R_{BL}$/$R_{WL}$ reduction, indicating that even the regular SRAM array cannot fully escape the local interconnect penalties of 4-tier stacking.

## V. NanoStack CFFET for Future Scaling

The preceding RO and parasitic evaluations revealed bottom-transistor access through supervias and VHV routing as the dominant source of CFFET performance penalty. Inspired by the Nanostack concept in [17], a **NanoStack CFFET** (Fig. 16) was proposed, in which FS and BS CFETs' actives are separately patterned and offset along the gate direction (Y-axis), opening direct M0 access to bottom transistors and eliminating VHV connections, however, at the cost of additional FS/BS litho. overlay requirements. The thermal budget issue could be addressed by the multi-flipping strategy [18]. At the INVD2 based RO level, NanoStack CFFET outperforms FFET by 3.1% at iso-power @ 0.7 V without BEOL loading (Fig. 17(a)), owing to the improved source resistance and Miller capacitance (Fig. 17(b)). Compared with FFET, the NFET/PFET's source resistances change from + 84.8%/+ 216.1% in baseline CFFET to - 36.6%/+ 31.5% in NanoStack CFFET, enabled by VHV elimination and closer NFET-to-VSS distance (Fig. 17(c)). These results highlight architecture-level DTCO for parasitic optimization beyond straightforward multi-tier transistor stacking, albeit at the cost of greater process complexity.

## VI. Can Logic Stacks Simply as Memory?

As 3D-NAND in volume production and 3D-DRAM being actively developed, similar concepts of multi-tier transistors for logic seem natural and straightforward, creating the myth that multi-tier 3D stacking is universally applicable for all circuits. However, as summarized in Table 3, logic standard cells, in great contrast to memories, are generally constrained by irregular routing, limited signal sharing, arbitrary transistor accessibility, great self-heating and strong device-to-interconnect coupling, leading to unforeseeably complex local routing vertically when multi-stacked. This obviously induces great challenges in local interconnect processes, regardless of the strict vertically patterning requirement for multiple N/PFETs. In summary, additional transistor tiers provide benefit only when their process, access and physical-design penalties are simultaneously resolved through deeper DTCO.

## VII. Conclusion

This work evaluated whether the nominal density gain of 4-tier stacked transistors can be translated into realistic block-level PPA benefits, with CFFET identified as the best stacking scheme. Despite the 40% nominal cell-height reduction, CFFET achieves only a 12.8% routed-area reduction while suffering 10.9% lower frequency than 2-tier-transistor FFET. This loss is rooted in two coupled constraints: additional MOL/VHV access increases parasitic RC and routing congestion, while 4-transistor cell building blocks reduce transistor utilization. Cell stitching recovers the area loss partially, whereas NanoStack CFFET gains RO performance with additional patterning and overlay complexity. This study unveils the undesired fact that the straightforward multi-stacking strategy may not be applicable for logic circuits and further serious investigation is needed.

## Acknowledgment

This work was supported in part by the National Key R&D Program of China under Grant 2023YFB4402201; in part by the NSFC under Grant 92464206; and in part by GJ Program under Grant 202504841098.

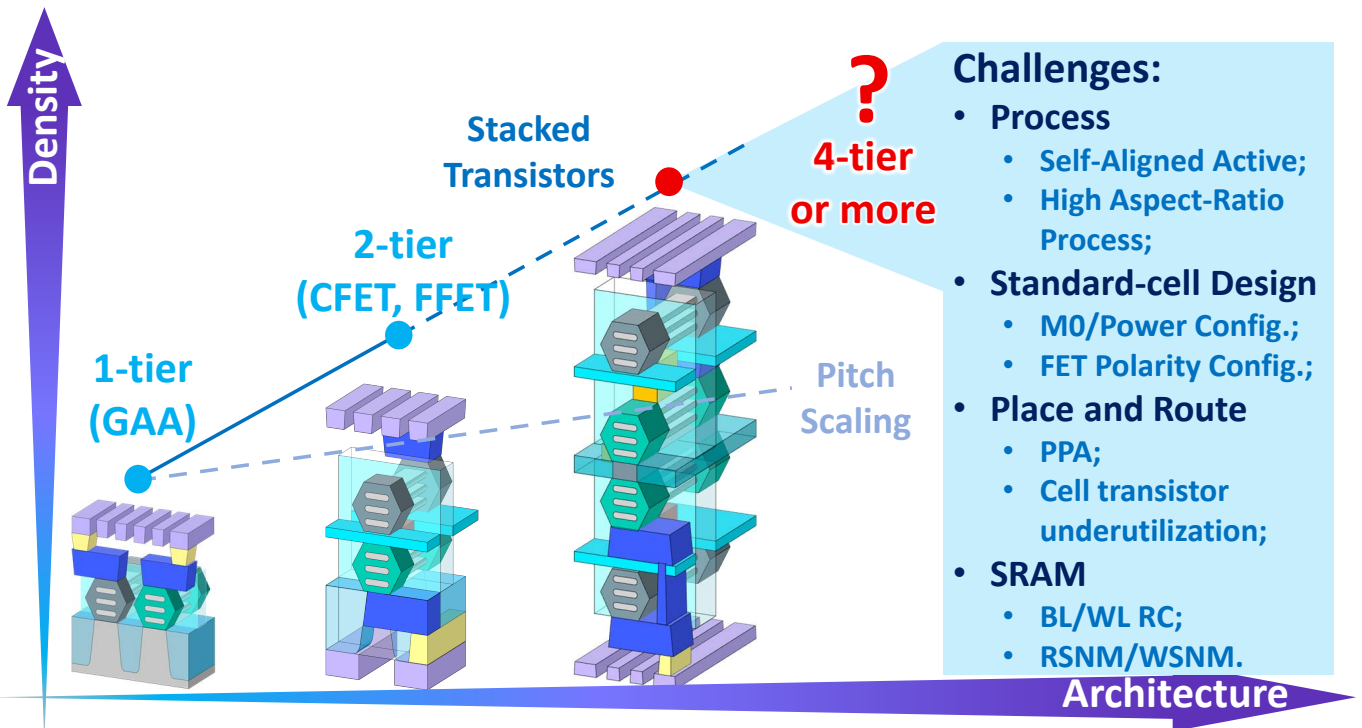

**Fig. 1** Roadmap of representative multi-tier transistor-stacking architectures and the emerging challenges that may limit multi-tier vertical stacked scaling.

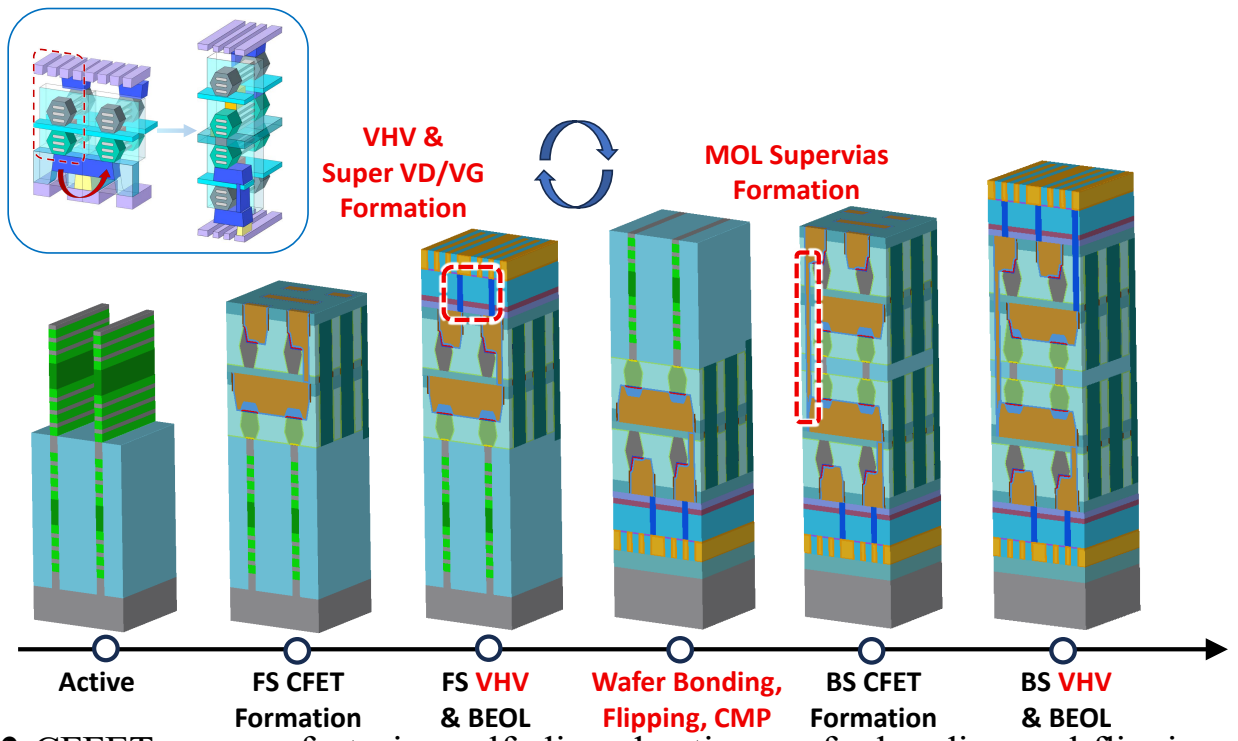

**Fig. 2** CFFET process featuring self-aligned active, wafer bonding and flipping, CFET on FS/BS of wafer and vertical access through VHV and supervias.

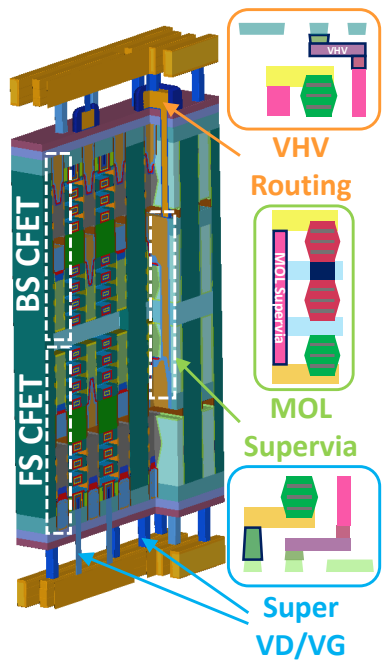

**Fig. 3** 3D structure of CFFET highlighting vertical access.

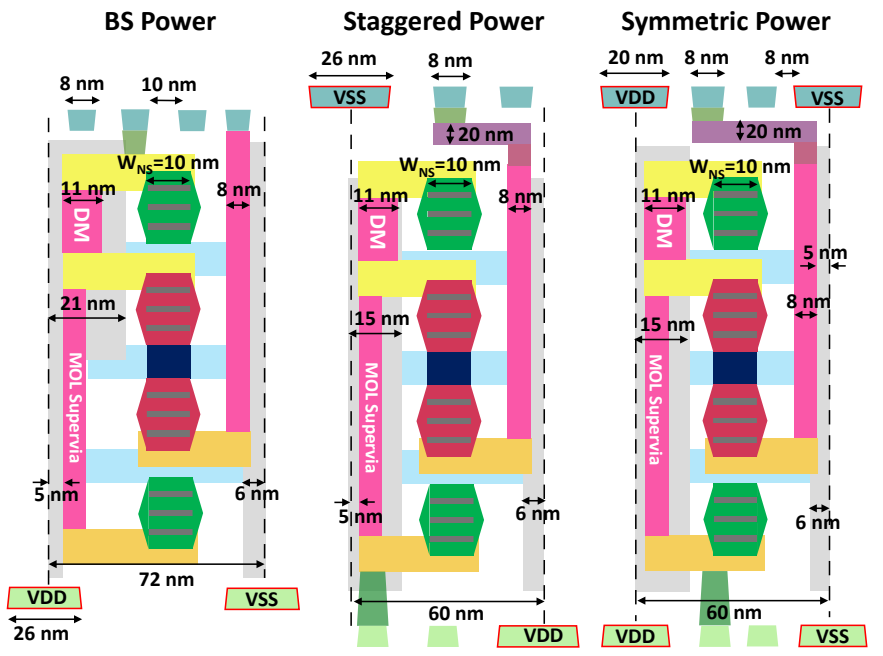

**Fig. 4** Design rules for three CFFET power configurations: BS Power, Staggered Power and Symmetric Power.

| Device | NSFET |
|---|---|
| $L_g$(nm) | 15 |
| $L_{gate\ metal}$(nm) | 11.8 |
| $N_{NS}$ | 3 |
| $W_{NS}$(nm) | 10 |
| $T_{NS}$(nm) | 5 |
| Gate Spacer Thk.(nm) | 5.5 |
| Inner Spacer Thk.(nm) | 6.5 |
| EOT(nm) | 0.83 |
| $I_{off}$(nA) | 2 |
| CPP(nm) | 45 |

**Table 1** Device assumptions for FFET and CFFET.

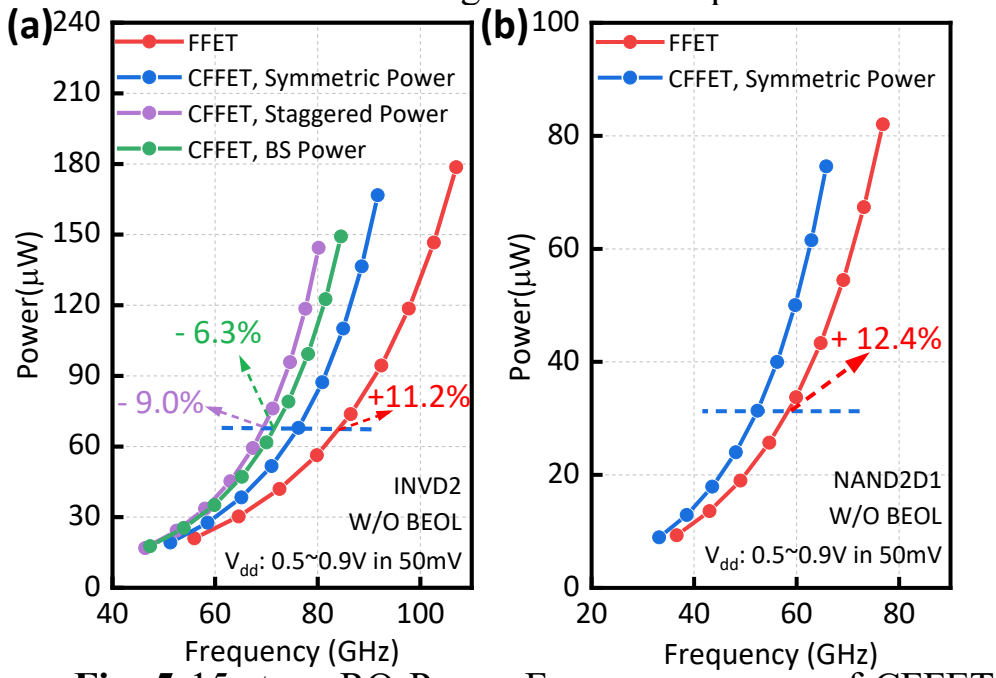

**Fig. 5** 15-stage RO Power-Frequency curves of CFFET benchmarked with FFET. (a) INVD2. (b) NAND2D1. FFET outperforms all power configurations of CFFET.

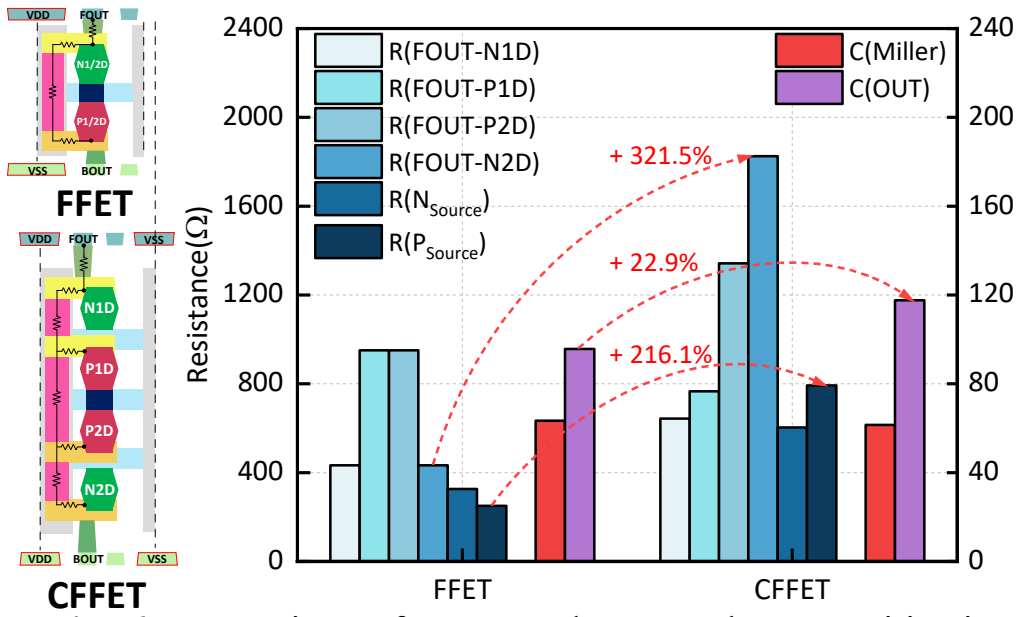

**Fig. 6** Comparison of FFET and CFFET key parasitics in INVD2. CFFET has larger parastics than FFET, especially in output/source resistance and output capacitance.

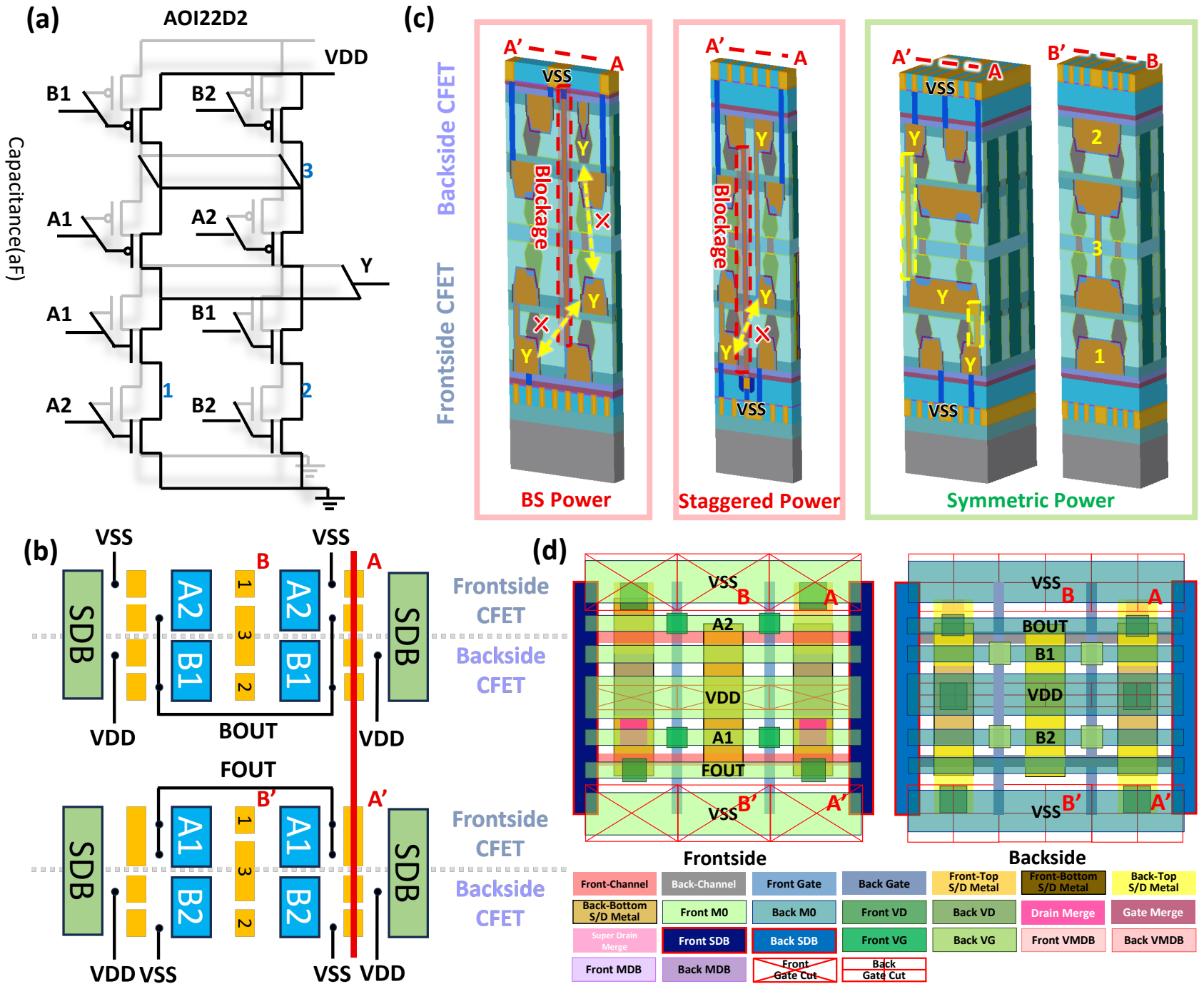

**Fig. 7** Layout design on three power configurations of CFFET. (a) Schematic of AOI22D2. (b) Schematic cross section of the 2-row AOI22D2 along the z-direction. (c) Cross section comparison of three power configurations of CFFET. The BS Power and Staggered Power have difficulty in MOL connection due to power supervias' blockage. (d) Layout of AOI22D2 for Symmetric Power.

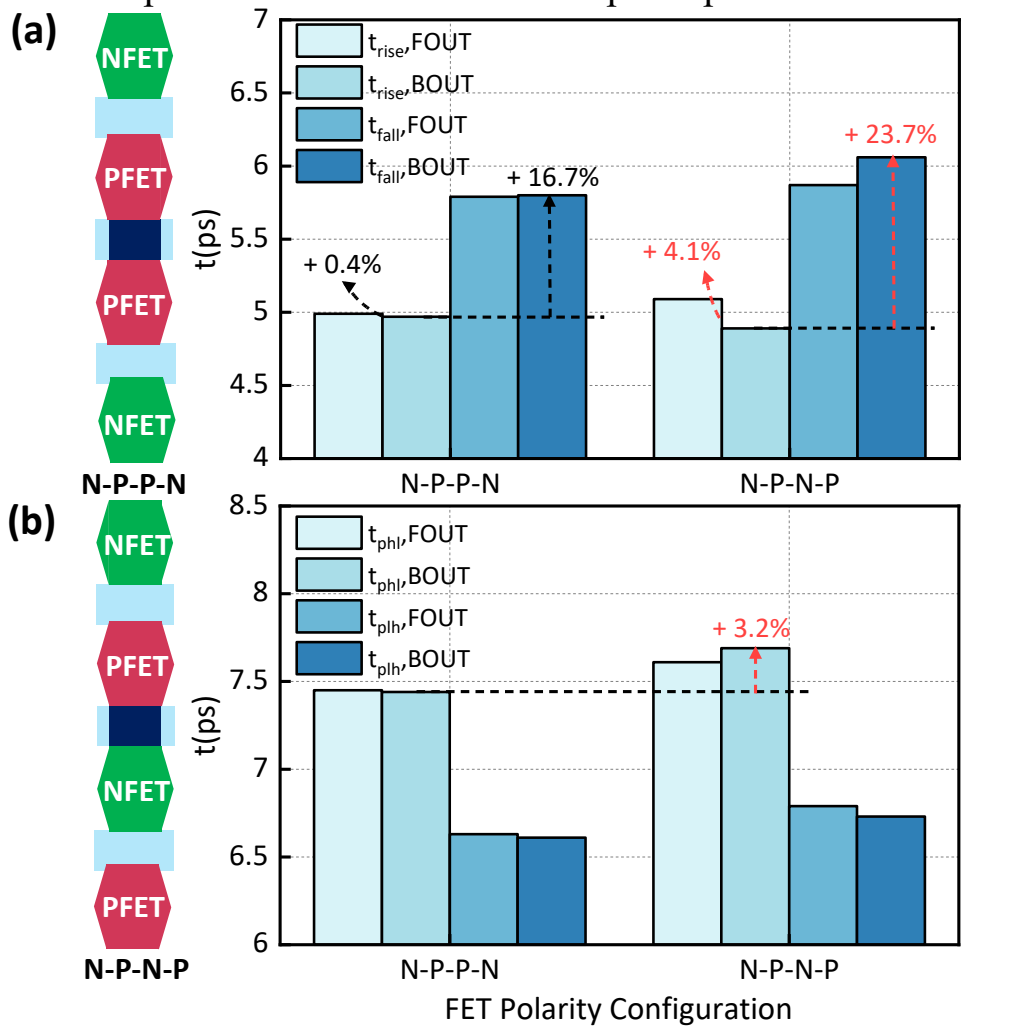

**Fig. 8** Comparison of output (a) transition time and (b) delay time for single-stage inverter with N-P-P-N and N-P-N-P FET polarity configurations.

| Architecture | Source Resistance | Performance | M0 Signal Tracks | Dependence to Edge Signal Track | Symmetric | MOL Routability | Cell Height per transistor |
|---|---|---|---|---|---|---|---|
| CFFET, Symmetric Power | Low | High | 4 (4 transistors) | Balanced | Highest | High | Lowest |
| CFFET, Staggered Power | Medium | Medium | 4 (4 transistors) | Balanced | Medium | Medium | Middle |
| CFFET, BS Power | High | Low | 4 (4 transistors) | High (Bottom gate, supervias) | Low | Low | High |
| FFET | Lowest | Highest | 4 (2 transistors) | Balanced | High | Highest | Highest |

**Table 2** Comparison of three power configurations of CFFET benchmarked with FFET.

**(a)**

| Layer | M0 | M1 | M2 | M3 | M4 | M5 | M6 |
|---|---|---|---|---|---|---|---|
| Pitch(nm) | 16 | 30 | 24 | 32 | 35 | 35 | 42 |
| Layer | M7 | M8 | M9 | M10 | M11 | M12 | M13 |
| Pitch(nm) | 76 | 76 | 76 | 76 | 76 | 76 | 126 |

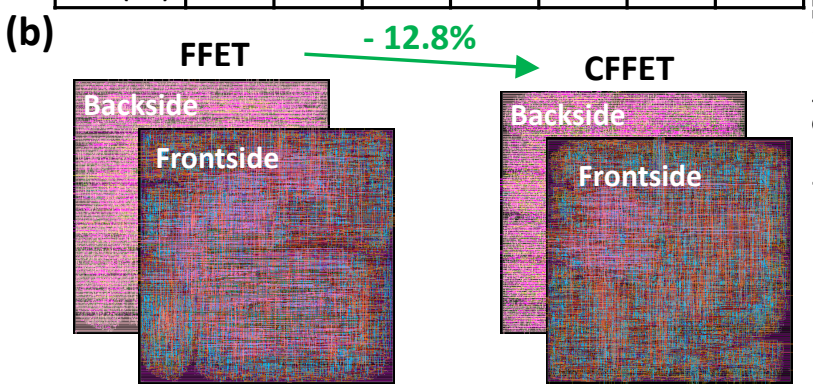

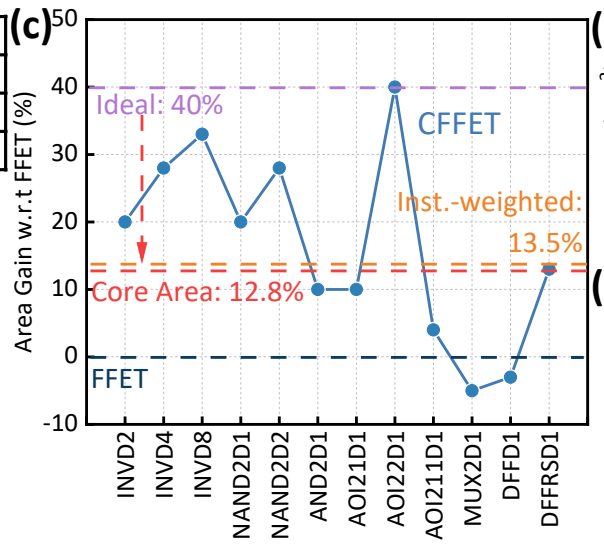

**Fig. 9** Physical implementation on Ibex RISC-V core. (a) BEOL metal design rule, shared between FS and BS. (b) Dual-sided layouts on FFET and CFFET. (c) Cell area comparison. The core area gain is much lower than ideal due to MOL congestion and transistor underutilization in CFFET standard-cell design.

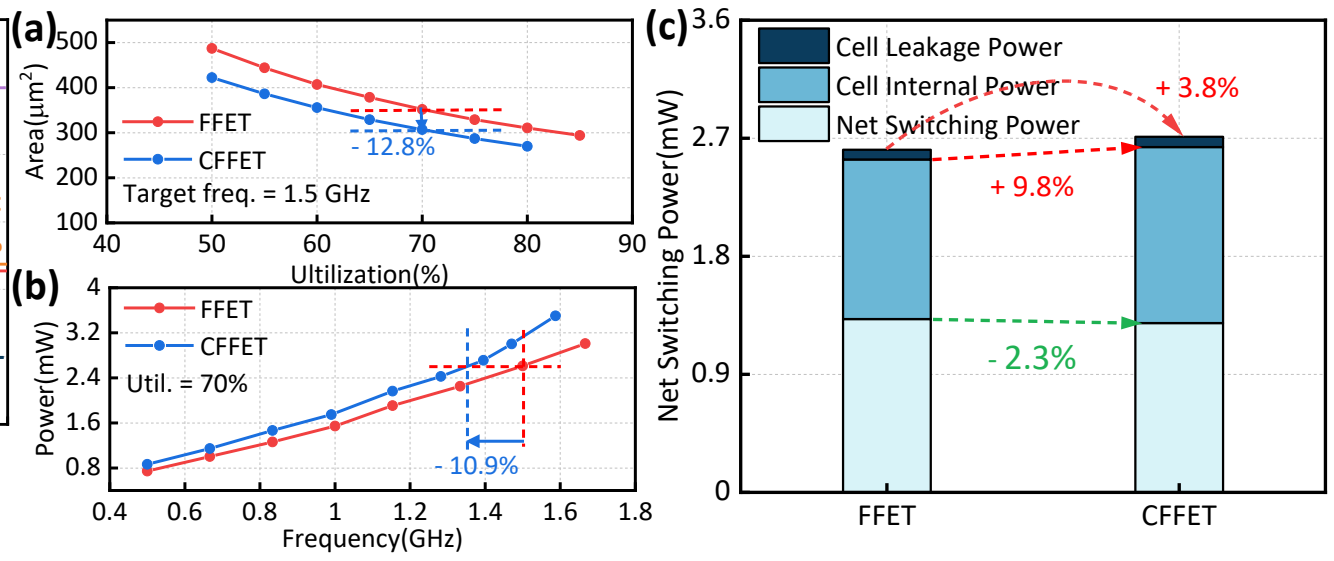

**Fig. 10** Block-level PPA results of FFET and CFFET Ibex cores: (a) Area-utilization plot at target frequency = 1.5 GHz. (b) Power-frequency plot at utilization = 70%. (c) Power breakdown of FFET and CFFET at 0.7 V, utilization = 70% and target frequency = 1.5 GHz.

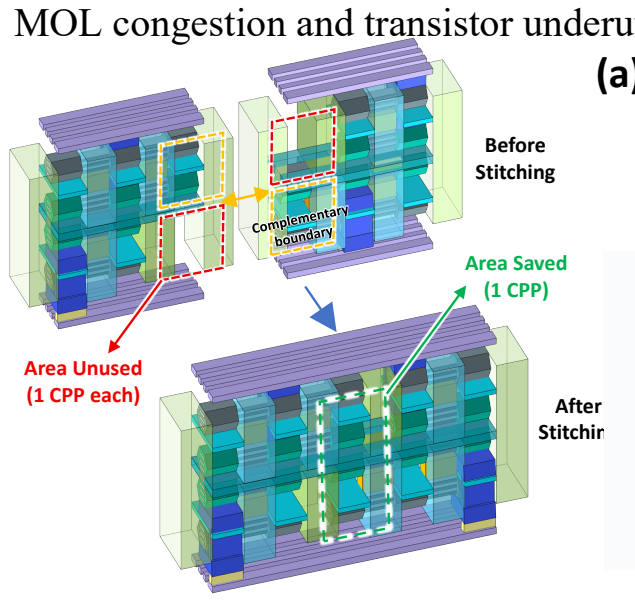

**Fig. 11** Schematic of Tetris-like cell stitching. Both cells have unused N/P pair next to cell boundary.

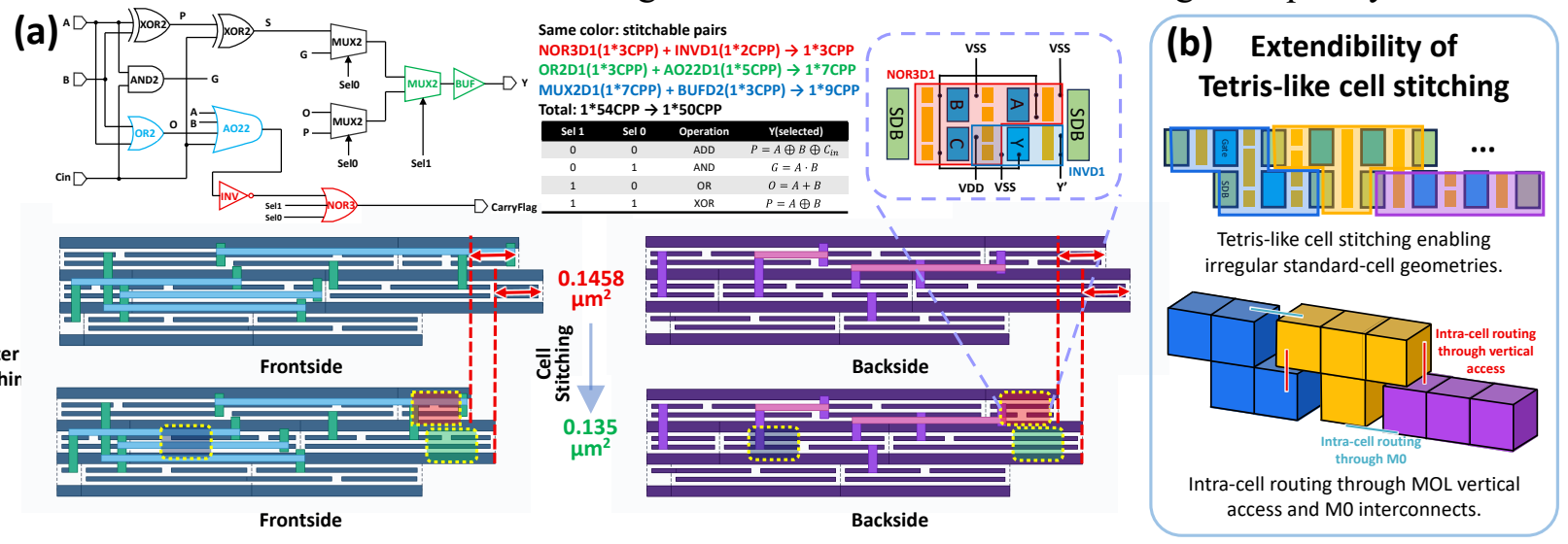

**Fig. 12** (a) Circuit schematic and functions of the illustrative 1-bit ALU, its FS/BS layouts before and after stitching with the stitching pairs highlighted, and a detailed NOR3D1–INVD1 stitching example. (b) Extendibility of Tetris-like cell stitching strategy.

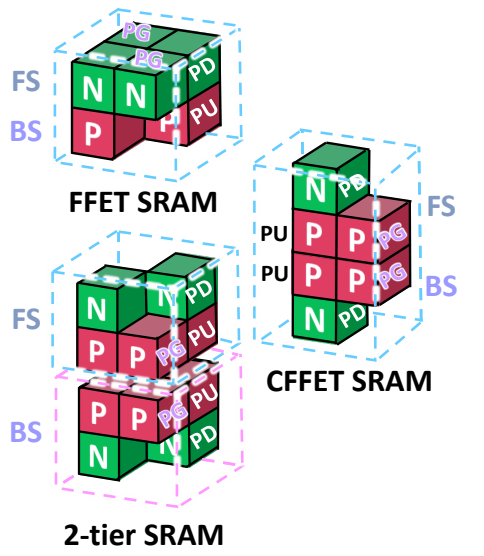

**Fig. 13** Schematic of FFET SRAM, 2-tier SRAM and CFFET SRAM.

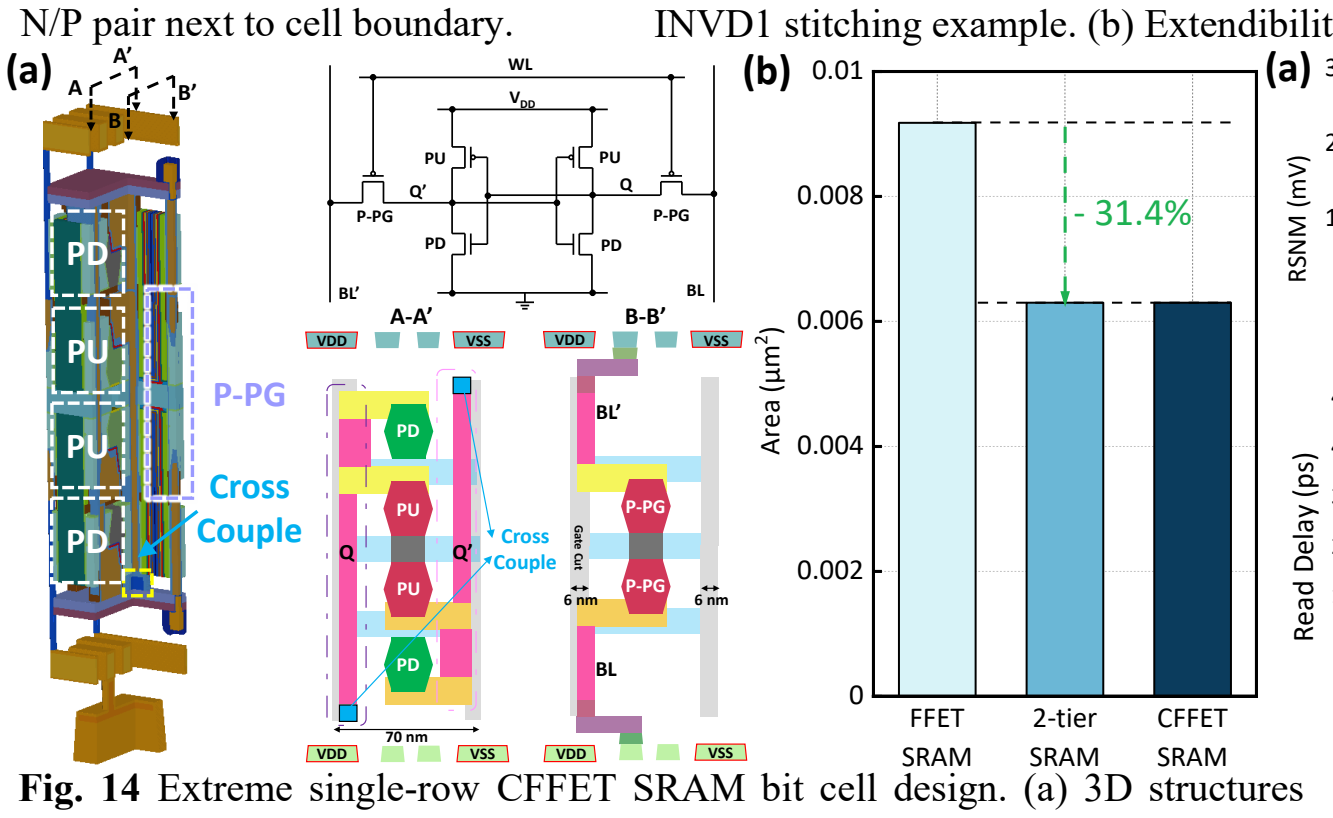

**Fig. 14** Extreme single-row CFFET SRAM bit cell design. (a) 3D structures and schematics in S/D cross sections. (b) Area scaling of CFFET SRAM and 2-tier SRAM design compared with FFET.

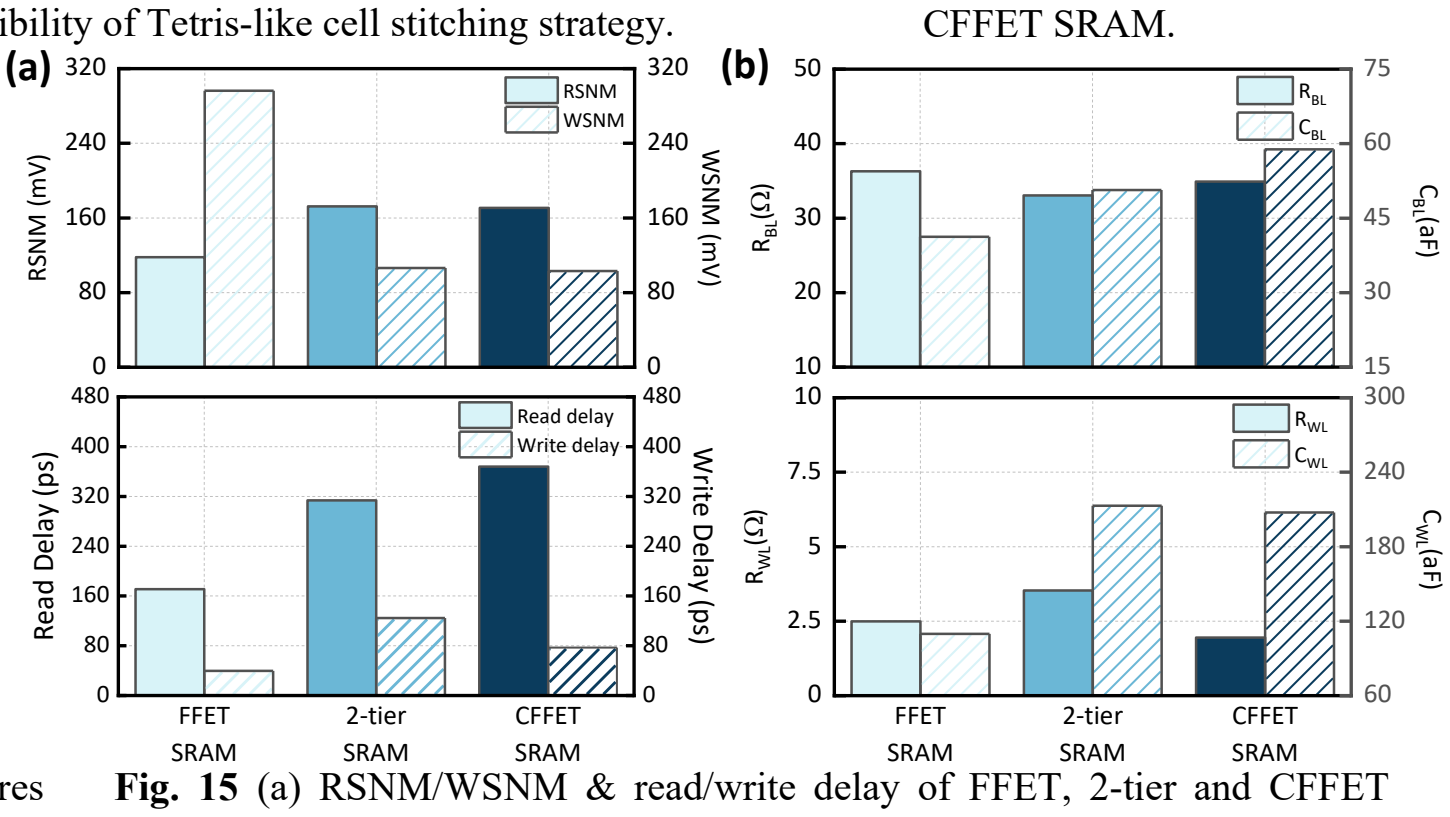

**Fig. 15** (a) RSNM/WSNM & read/write delay of FFET, 2-tier and CFFET SRAM bit cell. (b) WL & BL R&C per bit cell. Increased parasitics in 2-tier and CFFET SRAM leads to worse read/write performance than FFET.

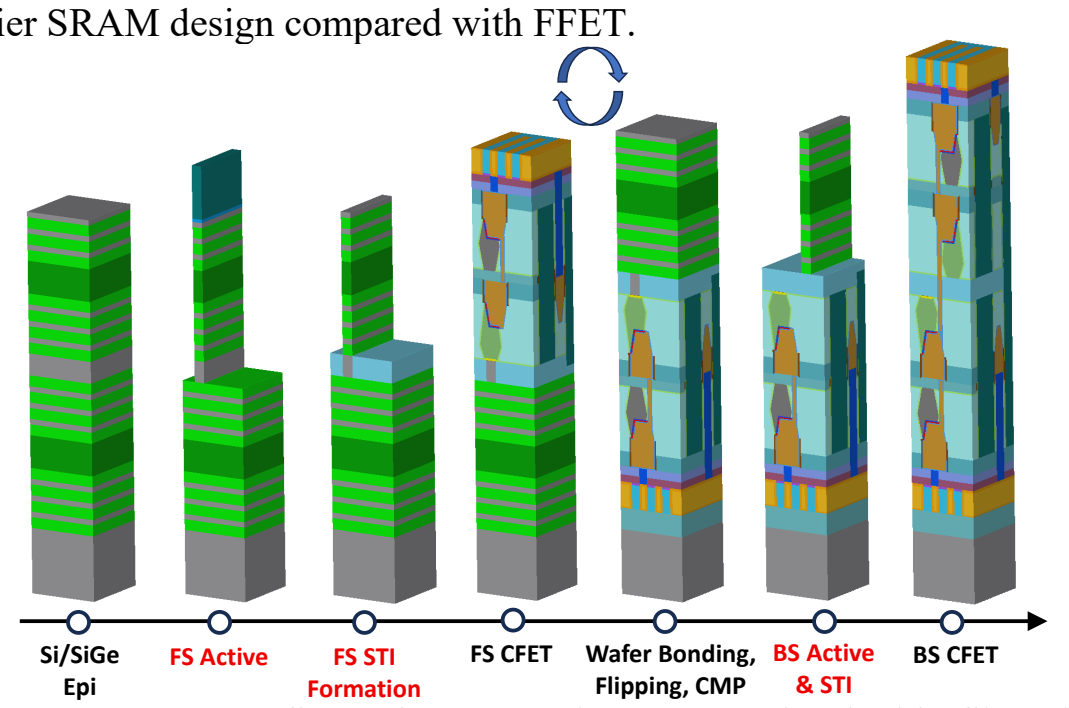

**Fig. 16** Process flow of NanoStack CFFET. The double-flipped process could also be implemented for thermal budget issue.

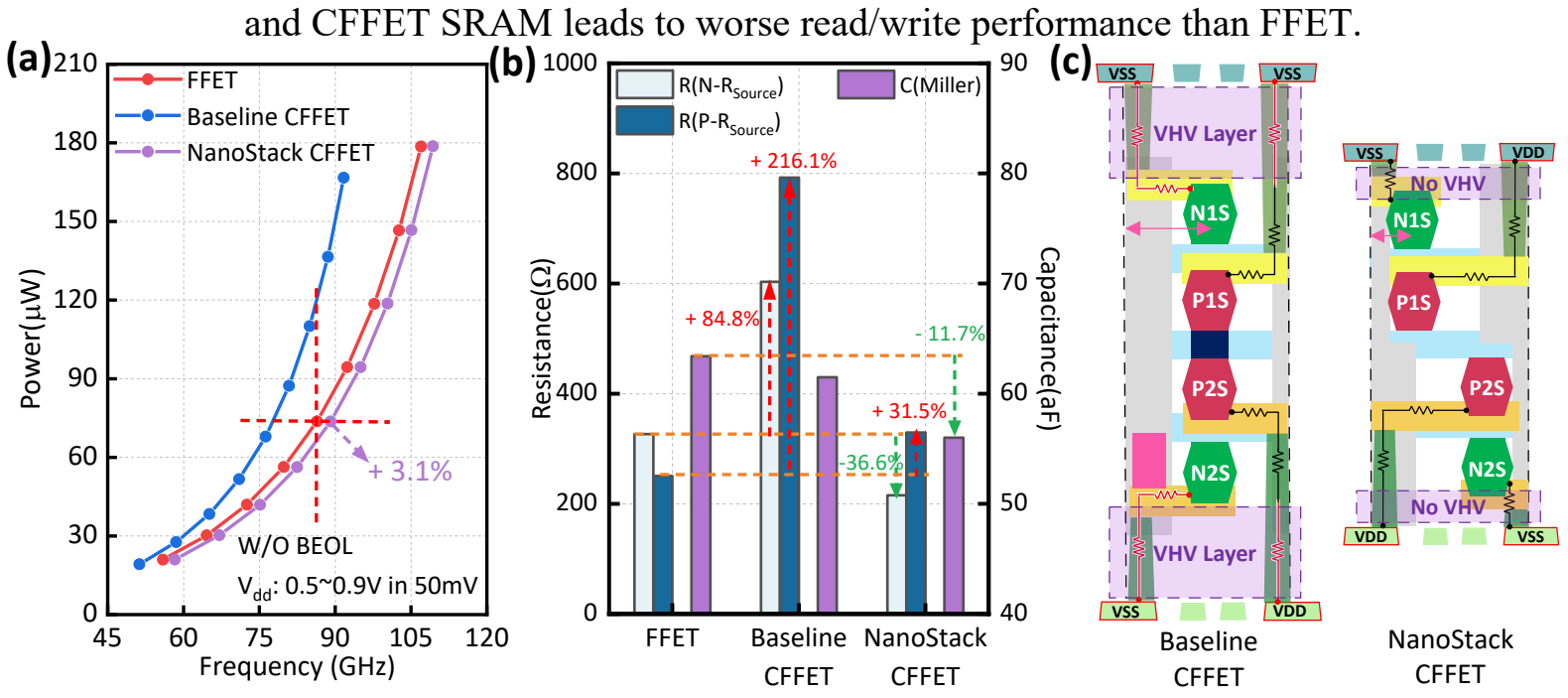

**Fig. 17** (a) Power-frequency of the 15-stage INVD2-based RO comparing FFET, Baseline and NanoStack CFFET. (b) Parasitics in INVD2. The frequency gain of Nanostack CFFET comes from smaller source resistance (as shown in (c)) and Miller capacitance.

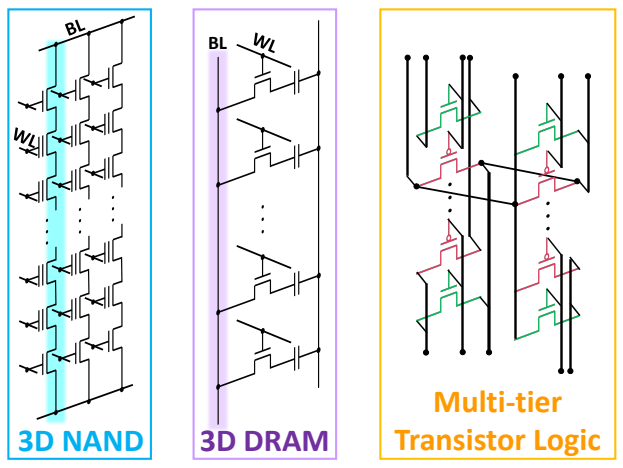

| Category | Stacking Architecture | Scaling Metrics | Key Metrics | Vertical Patterning | Connectivity | Structure | Vertical Path | Thermal Behavior | Defect Tolerance |
|---|---|---|---|---|---|---|---|---|---|
| NAND | 3D NAND | Bit density | Read/write latency, throughput, endurance | No needed | Shared WL/BL | Regular strings | Access path | Low | ECC / bad blocks |
| DRAM | 3D DRAM | Bit density | Access latency, bandwidth, retention | No needed | Array–periphery | Regular bit-cells | Sensing path | Low | ECC / redundancy |
| Logic | Multi-tier transistor | Post P&R PPA | Delay, frequency, power, energy efficiency | Separate N/P FETs needed | Arbitrary S/D/G access | Irregular logic connection, no simple replication | Timing-critical transistor-access path | Dense switching / clocks | Limited repairability |

**Table 3** Comparison of 3D NAND, 3D DRAM, and multi-tier logic transistors, highlighting the limits of multi-tier stacked transistors for logic.